\documentclass[a4paper,fleqn]{cas-dc}

\newcommand{\pd}[2]{\frac{\partial{#1}}{\partial{#2}}}

\newcommand{\od}[2]{\frac{\mathrm{d}{#1}}{\mathrm{d}{#2}}}

\usepackage[numbers,sort]{natbib}

\def\tsc#1{\csdef{#1}{\textsc{\lowercase{#1}}\xspace}}
\tsc{WGM}
\tsc{QE}

\begin{document}
\let\WriteBookmarks\relax
\def\floatpagepagefraction{1}
\def\textpagefraction{.001}

\shorttitle{}    

\shortauthors{}  

\title [mode = title]{Numerical study of the grain-growth-induced non-parabolic kinetics of a solid-state reaction}



%
\author[1]{Ya. A. Nikiforov}

\cormark[1]




\credit{Conceptualization, Methodology, Software, Formal analysis, Writing - Original Draft, Visualization}

\affiliation[1]{organization={Institute of Solid State Chemistry and Mechanochemistry SB RAS},
            addressline={18 Kutateladze st.}, 
            city={Novosibirsk},
            postcode={630090}, 
            country={Russia}}

\author[1]{S. A. Chizhik}




\credit{Methodology, Writing --- Review and Editing, Supervision}


\author[1]{N. I. Baklanova}

\credit{Writing --- Review and Editing, Supervision, Resources, Funding acquisition}
\cortext[1]{Corresponding author}



\begin{abstract}
	Non-parabolic growth of reaction layers is frequently observed
	in solid-state diffusion couples, but the relationship between
	microstructural evolution and the resulting kinetics remains
	incompletely understood. This effect is commonly attributed to grain growth,
	which can progressively reduce the contribution of fast grain-boundary 
	diffusion, causing the effective diffusivity of a polycrystalline
	product layer to evolve during reaction. This work  develops a
	moving-boundary diffusion model to describe product-layer growth
	while accounting for the local grain-growth history of the
	continuously formed product, as well as both bulk and grain-boundary
	diffusion. Numerical simulations show that the reaction can pass
	through three distinct kinetic regimes:
	an initial approximately parabolic regime dominated by grain-boundary
	diffusion, a transient sub-parabolic regime associated with strong
	spatial variation of the effective diffusivity, and a subsequent
	approximately parabolic regime dominated by bulk diffusion. 
	The magnitude and duration of the sub-parabolic regime depend
	on the relative grain-boundary and bulk diffusivities and on
	the kinetics of grain growth. The instantaneous growth exponent
	therefore evolves continuously and does not represent a unique
	kinetic constant. Nevertheless, fitting simulated layer-thickness
	data over finite experimental time intervals produces well-defined
	apparent exponents, demonstrating how a transient process can appear
	to obey a single power law.
\end{abstract}


\begin{keywords}
Solid-state reaction \sep 
Moving-boundary problem \sep
Computational modeling \sep
Grain growth \sep
Microstructural evolution \sep
Grain-boundary diffusion \sep
Non-parabolic kinetics\sep
\end{keywords}

\maketitle

\section{Introduction}\label{sec:intro}
Solid-state reactions between dissimilar materials commonly involve the formation
of one or more product phases at an initially sharp interface. Once a continuous
product layer has formed, its subsequent growth is frequently controlled by the
transport of one or more species through the product layer. Diffusion couples
therefore provide a useful framework for investigating reaction kinetics and
for relating it to the underlying transport processes \cite{Wagner1969,Kajihara2004,Mehrer2007}.
Under conditions for which the effective diffusivity of the product layer
can be regarded as constant, diffusion-controlled growth is described
by a parabolic law,
\begin{equation*}
\ell^2 \propto t,
\end{equation*}
where $\ell$ is the product-layer thickness and $t$ is time. This relationship
forms the basis of many classical treatments of reactive diffusion and is widely
used to characterize the kinetics of intermetallic and other solid-state reaction
layers \cite{Wagner1969,Kajihara2004,Mehrer2007}.

Experimental studies, however, have demonstrated that reaction-layer growth
does not always follow a simple parabolic law. Sub-parabolic growth,
conventionally expressed as
\begin{equation*}
\ell \propto t^\nu, \quad \nu < \frac{1}{2},
\end{equation*}
has been reported for a broad range of reactive-diffusion systems,
including intermetallic layers formed in such systems as Ni--Sn \cite{Taneja2018}, Pd--Sn \cite{Sakama2009}, Ag--Sn \cite{Li2010}, Al--Ti \cite{Zhang2019}, Al--Ni \cite{Azimi2019}, Ag--Zn \cite{Oh2024}, Cu--Zn \cite{Liu2020}, Fe--Al \cite{Xu2018}.
These observations indicate that departures from the classical parabolic law
are not restricted to a particular material system or processing configuration.
Such behavior is interpreted in terms of grain-boundary diffusion coupled with grain
growth, because coarsening reduces the density of fast diffusion paths available
for transport \cite{Schaefer1998,Ghosh2000,Wang2015,Breach2016}. In particular,
grain boundaries may provide diffusion paths with diffusivities considerably
higher than those associated with lattice transport. Consequently,
the effective diffusivity of a polycrystalline product layer depends
not only on temperature and composition but also on its evolving grain structure.

A number of analytical and numerical models have been proposed to establish
a quantitative relationship between microstructural evolution and reaction-layer
growth. An early and influential treatment was proposed by Schaefer et al. \cite{Schaefer1998}
for the growth of intermetallic compounds between  Cu and Sn–Pb solder.
Their model considered grain-boundary diffusion as the predominant transport
mechanism and explicitly incorporated the geometrical consequences of the
evolution of the grain structure. Importantly, the analysis demonstrated
that grain coarsening can modify the interdiffusion kinetics because
the density of high-diffusivity grain-boundary paths decreases as
the characteristic grain size increases. Although developed for a solid–liquid
soldering reaction, the underlying transport argument is not specific
to the presence of a liquid phase and provides an important basis
for relating grain-growth kinetics to non-parabolic product-layer growth.
A related analysis was subsequently presented by Ghosh \cite{Ghosh2000}
for the coarsening and growth of Ni\textsubscript{3}Sn\textsubscript{4}
during the reaction of liquid
solder with solid Ni/Pd layers. The study explicitly distinguished
the thickening and lateral coarsening kinetics of the intermetallic
morphology and found non-parabolic time dependences for both processes. 
These studies established 
an important conceptual link between microstructural coarsening
and deviations from the parabolic law, even though the specific
geometries and reaction configurations differ from those encountered
in solid-state diffusion couples.

Subsequent work extended this concept to solid-state interdiffusion
and multilayer intermetallic growth. Numerical study by Wang et al. \cite{Wang2015}
incorporated both grain-boundary and lattice diffusion into a model
for multiphase intermetallic growth and demonstrated that the relative
contribution of the two transport mechanisms changes systematically with grain size.
Xu et al. \cite{Xu2018} subsequently combined measurements of intermetallic
grain structure with a diffusion model for Fe--Al couples and showed that
grain growth and grain-boundary diffusion are important in determining
the observed thickening rates. These studies emphasize that a single,
time-independent interdiffusion coefficient may be inadequate when
the microstructure evolves significantly during the reaction.

A related analytical approach was developed by Nikiforov et al. \cite{Nikiforov2025},
who recently considered the effect of grain growth on the kinetics of the Ir–ZrC
solid-state reaction. In their model, grain coarsening produces a time-dependent
interdiffusion coefficient and leads to non-parabolic product-layer growth.
Under the assumptions of their asymptotic analytical treatment, a direct
relationship was obtained between the exponent describing grain growth and
that describing reaction-layer growth. The model was supported by experimental
measurements of both product-layer thickness and intermetallic
grain size \cite{Nikiforov2025}. This work provides an important theoretical
demonstration that the evolution of the product microstructure can itself
generate non-parabolic reaction kinetics.

Despite these advances, the existing analytical descriptions generally 
rely on simplifying assumptions concerning the dominant diffusion mechanism,
the initial grain size, or the relationship between grain size and product-layer
thickness. In particular, when the product layer grows continuously,
different portions of the layer are formed at different times and therefore
experience grain growth for different durations. The effective diffusivity
is consequently not merely a function of the total reaction time but varies
spatially according to the local microstructural history. A treatment that
simultaneously retains finite bulk diffusion, grain-boundary diffusion,
finite initial grain size, and this spatially varying grain-growth history 
is therefore required to describe the full transient evolution of the product layer.

The present work addresses this problem by considering a moving-boundary 
diffusion model in which the effective diffusion coefficient depends
explicitly on the local grain-growth history. The product layer is treated
as a continuously growing domain, while the local grain size is determined
by the time elapsed since the corresponding material element entered the
product phase. This formulation consequently gives rise to a diffusion
coefficient $D(x,t)$ that varies both spatially and temporally.
Both grain-boundary and bulk diffusion are retained, allowing the relative
importance of the two transport mechanisms to evolve naturally as
the product layer grows.

Rather than assuming a priori that the reaction-layer thickness follows 
a single power law, the present study evaluates the instantaneous growth
exponent and its evolution with time. The numerical results reveal three
successive regimes: an initial approximately parabolic regime associated
with predominantly grain-boundary transport, a transient sub-parabolic
regime associated with strong spatial variation of the effective diffusivity,
and a subsequent approximately parabolic regime in which bulk diffusion 
becomes dominant. The duration and magnitude of the intermediate regime
depend on the relative grain-boundary and bulk diffusivities as well as
on the kinetics of grain growth. Furthermore, because experimentally
determined growth exponents are generally obtained by fitting finite
sets of thickness measurements, the distinction between an instantaneous
kinetic exponent and an apparent exponent obtained from a power-law fit
is explicitly considered. The simulations indicate that a finite observation
interval can yield a well-defined apparent exponent even when the underlying
instantaneous exponent evolves continuously with time.

The objective of this work is therefore to establish how grain-growth-induced 
evolution of the effective diffusivity modifies the transient kinetics
of a diffusion-controlled solid-state reaction. Particular attention
is given to the transition between grain-boundary- and bulk-diffusion-dominated
regimes, the conditions under which sub-parabolic behavior emerges, and
the extent to which a finite experimental data set can be represented by
an apparently constant power-law exponent. Beyond grain growth itself,
this framework also provides a basis for considering other microstructural
processes that modify the effective diffusivity during evolution of the
product layer.

\section{Theoretical model}\label{sec:theory}

\subsection{Physical formulation}
\begin{figure}
  \centering
    \includegraphics{"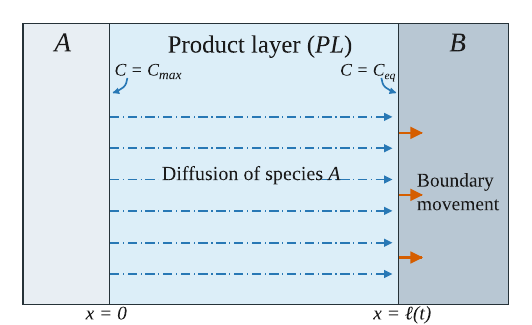"}
    \caption{Schematic representation of a diffusion couple with reaction
	    of boundary PL/B and diffusion of species A. Interface reaction
	    is assumed much more rapid than diffusion, so the concentration
    at two boundaries remain constant.}\label{fig1}
\end{figure}

Consider a diffusion couple (see Figure~\ref{fig1}) consisting of two phases,
denoted \textit{A} and \textit{B}, that react to form a product layer (\textit{PL}).
The geometry is one-dimensional, with $x=0$ denoting the \textit{A}/\textit{PL}
interface and $x=\ell(t)$ the moving \textit{PL}/\textit{B} reaction front.
Species \textit{A} diffuses through the product layer and is consumed 
at the moving reaction front. The reaction at the \textit{PL}/\textit{B}
interface is assumed to be sufficiently fast that the concentration
at this interface remains at its local equilibrium value. Likewise,
the concentration at the stationary \textit{A}/\textit{PL} interface is constant.
Under these assumptions, the reaction-layer growth is controlled by the diffusive
flux through the product layer.

The concentration $C(x,t)$ therefore satisfies Fick's second law
in a time-dependent domain,
\begin{equation}
	\pd{C}{t} = \pd{}{x}\left[D(x, t) \pd{C}{x}\right], \quad 0\leq x\leq \ell(t), 
	\label{eq:fick_eq_d_const}\\
\end{equation}
where $D(x,t)$ is the effective interdiffusion coefficient in the product layer.
The movement of the reaction front is determined by the flux arriving at $x=\ell(t)$,
\begin{equation}
	\frac{1}{\tilde{V}_A}\od{\ell}{t} = -D \left( \pd{C}{x} \right) \bigg|_{x=\ell}, \label{eq:stef_bc}
\end{equation}
where $\tilde{V}_A$ is the volume of product phase formed per atom of \textit{A}
consumed at the reaction front.

The boundary conditions are
\begin{equation}
	C(0, t) = C_\text{max}, \quad C(\ell(t), t) = C_\text{eq}. \label{eq:first_bc}
\end{equation}

\subsection{Effective diffusivity of the polycrystalline product layer}
The product layer is assumed to be polycrystalline and the effective interdiffusion
coefficient is evaluated as a volume-weighted combination of the bulk
and grain-boundary contributions \cite{Mishin1997},
\begin{equation}
	D = g D_{GB} +  (1-g)D_B
	\label{eq:diff_eff}
\end{equation}
where $D_B$ and $D_{GB}$ are the bulk and grain-boundary diffusion coefficients,
respectively, and $g$ is an effective volume fraction available for grain
boundary diffusion and is given by the equation:
\begin{equation}
	g = \frac{q\delta}{L}
	\label{eq:grain_frac}
\end{equation}
where $L$ is an average grain size, $\delta$ is the grain boundary width and $q$ 
is a numerical factor accounting for grain shape.

Equation~\ref{eq:diff_eff} can therefore be written as
\begin{equation}
	D = D_B + (D_{GB} - D_B) \frac{q \delta}{L}.
	\label{eq:diff_eff_2}
\end{equation}

Thus, grain growth introduces an intrinsic time dependence into the effective
interdiffusion coefficient.

\subsection{Grain growth and microstructural age}
The average grain size evolves according to the power law \cite{Abbaschian2008,Molodov2013}:
\begin{equation}
	L^m - L_0^m = k \tau
	\label{eq:grain_growth}
\end{equation}
where $L_0$ is the average grain size at the moment when a material element
enters the product layer, $k$ is the grain-growth constant, $m$ is the grain-growth
exponent, and $\tau$ is the time elapsed since formation of that material element.

Such formulation is more convenient than expressing grain size directly
as a function of the global reaction time because different locations
within the product layer are formed at different times.
If $t^*(x)$ denotes the time at which the moving reaction front
passes coordinate $x$, the local microstructural age is
\begin{equation}
	\tau(x, t) = t - t^*(x). \label{eq:def_age}
\end{equation}

Thus, $\tau = 0$ at the moving reaction front, whereas material farther toward
the stationary interface has existed within the product layer for a longer time.
The age field therefore provides a convenient description of the local
grain-growth history without explicitly tracking the passage time $t^*(x)$, 
as was done earlier in \cite{Nikiforov2025}.

The grain size is consequently
\begin{equation}
L(\tau) = \left( L_0^m + k \tau \right)^{1/m}.
	\label{eq:grain_size}
\end{equation}
Then, substitution into Eq.~\ref{eq:diff_eff_2} gives
\begin{equation}
	D(\tau) = D_B + (D_{GB} - D_B) q \delta \left(L_0^{m} + k \tau\right)^{-1/m}.
	\label{eq:diff_time_ap}
\end{equation}
The important consequence is that the diffusivity is determined by the local age of the product layer rather than by the global reaction time. The transformation in
Eq.~\ref{eq:def_age} converts the history-dependent transport problem into
a coupled diffusion–advection problem involving the additional scalar field $\tau$.

The age field satisfies
\begin{equation}
	\pd{\tau}{t} = 1,
\end{equation}
at fixed laboratory coordinate $x$, together with the boundary condition
\begin{equation}
\tau(\ell(t), t) = 0.
\label{eq:tau_bc}
\end{equation}

Thus, the complete system of differential equations describing the problem can 
be written as
\begin{gather}
	\pd{C}{t} = \pd{}{x}\left[D(\tau) \pd{C}{x}\right], \quad 0\leq x\leq \ell(t) 
	\label{eq:diff_eq_final}
	\\
	\frac{1}{\tilde{V}_A}\od{\ell}{t} = -D(0) \left( \pd{C}{x} \right) \bigg|_{x=\ell(t)}
	\label{eq:boundary_eq_final}
	\\
	\pd{\tau}{t} = 1,
	\label{eq:age_eq_final}
\end{gather}
with Eqs.~\ref{eq:first_bc} and \ref{eq:tau_bc} as boundary conditions.

\subsection{Asymptotic behavior}

Consider the asymptotic limit corresponding to negligible bulk diffusion
and vanishing initial grain size, i.e., $D_B \rightarrow 0$ and $L_0 \rightarrow 0$.
This case was thouroughly examined by Nikiforov et al. \cite{Nikiforov2025}, so we 
state here only the essential results. Under these conditions, transport is
governed by grain-boundary diffusion, while the equation for grain size evolution
(Eq.~\ref{eq:grain_size}) transforms into
\begin{equation*}
L = (k / \tau)^{1/m}.
\end{equation*}

The resulting reaction-layer growth can be expressed in the form
\begin{equation}
\ell = \lambda t^{\nu},
\end{equation}
where the kinetic exponent is related to the grain-growth exponent according to
\begin{equation}
\nu=\frac{m-1}{2m}.
\label{eq:link_equation}
\end{equation}
For the grain-growth exponents considered in this study, the relation \ref{eq:link_equation} gives
\begin{align*}
m=2:\quad \nu=\frac14,
\\
m=3:\quad \nu=\frac13,
\\
m=4:\quad \nu=\frac38.
\end{align*}
These values correspond to the asymptotic exponents and provide useful reference
limits for assessing the numerical results obtained with finite $D_B$ and $L_0$.

\section{Numerical solution}\label{sec:num_scheme}
\subsection{Transformation to a fixed computational domain}
The moving boundary in Eqs.~\ref{eq:diff_eq_final}--\ref{eq:age_eq_final}
is inconvenient for direct numerical integration. We therefore introduce
the normalized spatial coordinate
\begin{equation}
\xi = \frac{x}{\ell(t)}, \qquad 0 \leq \xi \leq 1,
\end{equation}
which maps the time-dependent product layer onto a fixed computational domain.

The spatial derivative of a scalar field transforms according to
\begin{equation}
\left(\pd{}{x}\right)_{t}= \frac{1}{\ell(t)} \left(\pd{}{\xi}\right)_{t}.
\end{equation}

Because the coordinate transformation itself depends on time
\begin{equation}
	\left( \pd{\xi}{t} \right)_{x} = - \xi \frac{\dot{\ell}}{\ell},
\end{equation}
the time derivative also needs to be transformed according to
\begin{equation}
	\left( \pd{}{t} \right)_{x} = \left( \pd{}{t} \right)_{\xi} - \xi \frac{\dot{\ell}}{\ell} \pd{}{\xi}.
\end{equation}

Application of these transformations to Eq.~\ref{eq:diff_eq_final} gives 
\begin{equation}
	\pd{C}{t} = \frac{1}{\ell^2}\pd{}{\xi}\left[D(\tau) \pd{C}{\xi}\right] + \frac{\dot{\ell}}{\ell} \xi \pd{C}{\xi}, \quad 0\leq \xi \leq 1.
	\label{eq:diff_eq_transform}
\end{equation}
Equation on the moving boundary, Eq.~\ref{eq:boundary_eq_final}, becomes
\begin{equation}
	\od{\ell}{t} = -\frac{\tilde{V}_A D(0)}{\ell} \left( \pd{C}{\xi} \right) \bigg|_{\xi=1}.
	\label{eq:boundary_eq_transform}
\end{equation}
The age equation requires the same coordinate transformation. Hence, at fixed $\xi$
\begin{equation}
	\pd{\tau}{t} = 1 + \frac{\dot{\ell}}{\ell} \xi \pd{\tau}{\xi}.
	\label{eq:age_eq_transform}
\end{equation}
The boundary conditions now read
\begin{equation}
	C(0, t) = C_\text{max}, \quad C(1, t) = C_\text{eq},
\end{equation}
and 
\begin{equation}
\tau(1, t) = 0.
\end{equation}

\subsection{Dimensionless formulation}
The model contains several dimensional quantities whose individual values are not required to determine the qualitative kinetic behavior. We therefore introduce the initial grain size as the characteristic length, $L_0$, and the corresponding bulk-diffusion time, $t_0 = L_0^2 / D_B$.
The dimensionless variables are defined as
\begin{equation*}
t^\dagger = t / t_0, \quad \tau^\dagger = \tau / t_0, \quad \ell^\dagger = \ell / L_0,
\end{equation*}
and 
\begin{equation}
C^\dagger = \tilde{V}_{A} C.
\end{equation}

The choice of $L_0$ as the characteristic lenth gives 
\begin{equation*}
L_0^\dagger = 1,
\end{equation*}
while the choice of $t_0$ gives 
\begin{equation*}
D_B^\dagger = 1.
\end{equation*}

The grain-growth constant is transformed into dimensionless parameter
\begin{equation*}
k^\dagger = \frac{k L_0^2}{D_B},
\end{equation*}
and the dimensionless measure of grain-boundary transport is
\begin{equation}
	\mathcal{D}_{GB}^\dagger = \frac{q \delta }{L_0} \frac{D_{GB} - D_B}{D_B}.
\end{equation}
his parameter combines two physically distinct effects: the diffusivity ratio 
$(D_{GB} - D_B) / D_B$ and the geometrical fraction $q \delta / L_0$
associated with grain boundaries. Thus, $\mathcal{D}_{GB}^{\dagger}$ measures 
the maximum enhancement of the effective diffusivity relative to bulk diffusion at 
the initial grain size.

After omitting the daggers for clarity, the dimensionless effective diffusivity becomes
\begin{equation}
	D(\tau) = 1 + \mathcal{D}_{GB}\left[ 1 + k \tau \right]^{-1/m}.
	\label{eq:diffusivity_dimless}
\end{equation}

The governing equations are then 
\begin{gather}
	\pd{C}{t} = \frac{1}{\ell^2}\pd{}{\xi}\left(D(\tau) \pd{C}{\xi}\right) + \frac{\dot{\ell}}{\ell} \xi \pd{C}{\xi},
	\label{eq:diff_eq_dimless}
	\\
	\od{\ell}{t} = -\frac{D(0)}{\ell} \left( \pd{C}{\xi} \right) \bigg|_{\xi=1},
	\label{eq:boundary_eq_dimless}
	\\
	\pd{\tau}{t} = 1 + \frac{\dot{\ell}}{\ell} \xi \pd{\tau}{\xi}.
	\label{eq:age_eq_dimless}
\end{gather}

Thus, after non-dimensionalization, the model is governed by only three independent parameters: $\mathcal{D}_{GB}$, $k$, and $m$. $\mathcal{D}_{GB}$ controls the relative
importance of grain-boundary transport, $k$ controls the rate of microstructural
coarsening relative to bulk diffusion, and $m$ specifies the mode of grain growth.

The present calculations consider $m = 2, 3, 4$, representing normal grain growth
and increasingly retarded grain growth, respectively. The explored range of
$\mathcal{D}_{GB}$ is $1$--$10^3$, 
while $k$ is varied over several orders
of magnitude relative to $\mathcal{D}_{GB}$.


\subsection{Computational scheme}

The fixed computational domain $0 \leq \xi \leq 1$ is discretized using $M$
uniformly spaced nodes,
\begin{equation*}
\Delta \xi = \frac{1}{M - 1}.
\end{equation*}

Time integration is performed using adaptive time steps. Adaptivity is necessary
because the reaction-layer growth rate changes substantially during the simulation:
small time steps are required during the early stage, whereas larger steps can
be used after the growth rate decreases.

The time step is restricted by both diffusion and domain-expansion timescales.
We define
\begin{equation}
	\Delta t_\text{diff} = \alpha \frac{\ell^2}{D_\text{max}},
\end{equation}
and
\begin{equation}
	\Delta t_\text{adv} = \beta \frac{\ell}{\dot{\ell}},
\end{equation}
where parameters $\alpha$ and $\beta$ are chosen to maintain stability. The time step
$\Delta t$ used at each iteration is the minimum between $\Delta t_\text{diff}$ and $\Delta t_\text{adv}$.

Each time steps consists of three sequential updates. First,
eq.~\ref{eq:boundary_eq_dimless} is solved by the forward Euler scheme:
\begin{equation*}
	\ell^{n+1} = \ell^n +  \dot{\ell}^{n} \Delta t,
\end{equation*}
where 
\begin{equation}
\dot{\ell}^{n} = - \frac{D_{M-1}^{n}}{\ell^n}
	\frac{C_{M-1}^{n} - C_{M-2}^{n}}{\Delta \xi}.
\end{equation}

Next, the age-field equation, Eq.~\ref{eq:age_eq_dimless}, is advanced using
a semi-implicit Crank-Nicholson scheme. The discretization leads to a 
linear tridiagonal system of equations
\begin{equation}
	-S_i^{n+1} \tau_{i+1}^{n+1} + \tau_i^n + S_i^{n+1} \tau_{i-1}^{n+1}
	= \tau_i^n + \Delta t + S_i^n \left(\tau_{i+1}^{n} - \tau_{i-1}^{n} \right)
\end{equation}
where
\begin{equation}
	S_i^n = \frac{\Delta t}{4 \Delta \xi} \frac{\dot{\ell}^{n}}{\ell^n} \xi_i
	\quad \text{and} \quad
	S_i^{n+1} = \frac{\Delta t}{4 \Delta \xi} \frac{\dot{\ell}^{n}}{\ell^{n+1}} \xi_i .
	\label{eq:s_matrices}
\end{equation}
Here we adopt a lagged evaluation of the moving boundary velocity,
$\dot{\ell}^{n+1} \approx \dot{\ell}^{n}$. That forces us to use small
time step $\Delta t$ but greatly reduces the size of systems of linear 
equations to solve.

After updating $\tau$, the local diffusion coefficient is
evaluated from Eq.~\ref{eq:diffusivity_dimless},
\begin{equation*}
	D_i^{n+1} = 1 + \mathcal{D}_{GB} \left[ 1 + k \tau_i^{n+1} \right]^{-1/m}.
\end{equation*}

Finally, the concentration equation, Eq.~\ref{eq:diff_eq_dimless} is advanced
using a semi-implicit Crank-Nicholson scheme
\begin{equation}
	\begin{split}
	& - (A_i^{n+1} + S_i^{n+1}) C_{i+1}^{n+1} + (1 + A_i^{n+1} + B_i^{n+1}) C_i^{n+1}\\
	& + (S_i^{n+1} - B_i^{n+1}) C_{i-1}^{n+1} = (A_i^{n} + S_i^{n}) C_{i+1}^{n} \\
	&+ (1 - A_i^{n} - B_i^{n}) C_i^{n} + (B_i^{n}- S_i^{n}) C_{i-1}^{n}, 
	\end{split}
\end{equation}
where $S_i^n$ and $S_i^{n+1}$ are defined in eq.~\ref{eq:s_matrices}, and matrix elements $A_i^n$ and $B_i^n$ are:
\begin{align}
	A_i^n &= \frac{\Delta t}{2 (\ell^n \Delta \xi)^{2}} \frac{D_{i+1}^{n} + D_{i}^{n}}{2},\\
	B_i^n &= \frac{\Delta t}{2 (\ell^n \Delta \xi)^{2}} \frac{D_{i+1}^{n} + D_{i-1}^{n}}{2}.
\end{align}
Similar equations with $n$ changed to $n+1$ apply to $A_i^{n+1}$ and $B_i^{n+1}$.

This sequence is repeated until the prescribed final time is reached.

The numerical calculation starts from a finite product-layer thickness rather
than from $\ell(0) = 0$. The initial layer thickness is taken equal to the initial
grain size, which corresponds to $\ell(0) = 1$ in dimensionless units.

Also, a linear concentration profile is imposed initially, consistent with the assumption that diffusion is initially rapid and that the earliest stage of the physical reaction is influenced by the finite initial layer and interfacial reaction.

Accounting for this finite initial layer is important when comparing the numerical
solution with the idealized analytical solution for a product layer initially
of zero thickness. If the time is not shifted to account for the time
needed to produce this initial tickness $\ell(0)$, an artificial deviation
from the analytical solution appears. Consequently, the analysis of kinetic
exponents should be performed after the initial transient time has diminished 
with respect to the simulation time.

\subsection{Characterization of the growth kinetics}

The principal observable is the product-layer thickness $\ell(t)$. A power law
\begin{equation}
\ell(t) = \lambda t^{\nu},
\label{eq:power_law}
\end{equation}
is commonly used to characterize reaction-layer growth. For constant diffusivity,
diffusion-controlled growth gives $\nu=1/2$. However, Eq.~\ref{eq:power_law}
need not represent an exact solution of the present model because
the effective diffusivity evolves as a consequence of grain growth.

To distinguish an instantaneous kinetic behavior from an exponent obtained 
by fitting a finite data set to Eq.~\ref{eq:power_law}, we define
the instantaneous growth exponent as
\begin{equation}
\nu(t) = \frac{\mathrm{d} \log(\ell)}{\mathrm{d} \log(t)}.
\end{equation}
For an exact power law, $\nu(t)$ is constant. In the present problem,
variations in $\nu(t)$ provide a direct measure of the departure
from a single power-law description.

\subsection{Numerical representation of experimentally measured kinetics}

To assess how the calculated kinetics would be interpreted experimentally,
the numerically modelled $\ell(t)$ curves are also analyzed over finite time intervals.
A set of points is selected from a time interval $[t_1, t_2]$ which can be measured
experimentally in a laboratory (from $\sim 1$ to $\sim 100$~hours) and random
Gaussian perturbations corresponding to a 5\% measurement uncertainty
are applied to the layer thickness. The resulting data are transformed according to
\begin{equation}
	\log{\ell} = \log{\lambda} + \bar{\nu} \log{t},
	\label{eq:log_power_law}
\end{equation}
and the slope obtained by linear regression is taken as the apparent
exponent $\bar{\nu}$.

Repeated sampling is used to determine both the mean apparent exponent 
and its statistical variation. This procedure follows the experimental perspective:
the objective is not to assign a new physical exponent to the model,
but to determine can a well-defined exponent be inferred if the continuously
evolving kinetics were measured over a finite experimental time window.


\section{Results and discussion}\label{sec:rnd}

\subsection{Validation against the constant-diffusivity solution}

\begin{figure}
  \centering
    \includegraphics{"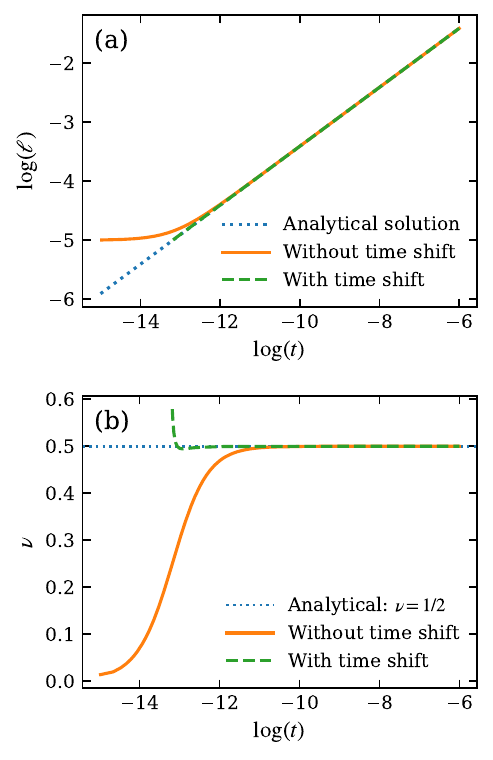"}
    \caption{Validation of the numerical model against the analytical solution for diffusion-controlled product-layer growth in the absence of grain growth.
	    (\textit{a}) Product layer thickness, $\ell$, as a function of time in 
    logarithmic coordinates;
    (\textit{b}) Corresponding instantaneous growth exponent $\nu$ as a 
    function of $\log(t)$.
    The numerical results are shown before and after applying the time shift associated with the finite initial product-layer thickness.}\label{fig2}
\end{figure}

As a first step, the numerical formulation was evaluated for the limiting case
in which grain growth is absent (k=0). Under these conditions, the effective
diffusion coefficient is spatially and temporally invariant, and the problem
reduces to the classical diffusion-controlled growth of a planar product layer.
The analytical solution is therefore expected to exhibit parabolic growth,
\begin{equation}
	\ell(t) = 2 \gamma \sqrt{Dt}, \label{eq:parabolic_growth}
\end{equation}
where $\gamma$ is determined from 
\begin{equation}
	\frac{C_\text{max} - C_\text{eq}}{\sqrt{\pi}} = \gamma \operatorname{erf}{(\gamma)} \exp{(\gamma^2)}.
\end{equation}

The corresponding concentration profile is given by the similarity solution
expressed in the normalized coordinate, $\xi=x/\ell$,
\begin{equation}
	C(\xi) = C_\text{max} - \frac{C_\text{max} - C_\text{eq}}{\operatorname{erf}(\gamma)} \operatorname{erf} (\gamma \xi).
\end{equation}

Figure~\ref{fig2}a presents the simulated curve $\ell(t)$ in logarithmic coordinates 
against the analytical solution. A mentioned earlier in Section~\ref{sec:num_scheme},
the numerical calculation starts from $\ell(0)=1$ and consequently the numerical time
origin does not coincide exactly with that of the analytical solution.
If this difference is neglected, an apparent deviation from parabolic growth
is observed during the initial stage of the calculation.
Introducing the corresponding time shift, estimated from Eq.~\ref{eq:parabolic_growth},
substantially improves the agreement between the numerical and analytical solutions.

The same effect is evident from the instantaneous growth exponent
shown on Fig.~\ref{fig2}b. For the analytical solution, $\nu=1/2$ at all times.
The numerical solution exhibits a pronounced deviation from this value
at the beginning of the simulation, particularly when the time shift is omitted.
The deviation progressively decreases as the simulation time becomes large
compared with the characteristic time required to establish the initial
product-layer thickness. Thus, the early portion of the numerical solution
should not be interpreted as intrinsic reaction kinetics.

\subsection{Evolution of the instantaneous growth exponent}

The influence of grain growth on reaction-layer kinetics was examined
by considering a finite grain-growth constant $k$. The calculated instantaneous
exponent, $\nu(t)$, provides a more informative description of the kinetics
than a single power-law exponent because the effective diffusivity evolves
continuously as the product-layer microstructure coarsens.

For relatively weak grain-boundary transport, represented by
$\mathcal{D}_{\mathrm{GB}} = 1$, the deviation from parabolic kinetics is small.
The growth exponent remains close to $1/2$ over most of the investigated time range,
although a shallow minimum develops at intermediate times.
This behavior originates from the spatially non-uniform evolution
of the diffusion coefficient. At early times, the product layer
consists predominantly of material with a relatively young microstructural age,
and the effective diffusivity is therefore close to its maximum value.
At sufficiently long times, most of the product layer has undergone
substantial grain coarsening, reducing the grain-boundary contribution
and bringing the effective diffusivity close to the bulk value.
In both limiting cases, the diffusivity is approximately uniform 
across the product layer and the growth approaches the parabolic regime.

The effect becomes substantially stronger as the relative contribution 
of grain-boundary diffusion is increased. For $\mathcal{D}_{\mathrm{GB}}=100$
and 1000 (see Fig.~\ref{fig4}), the minimum in $\nu(t)$ becomes pronounced,
demonstrating that grain growth can generate a substantial departure from
classical parabolic kinetics even though the reaction remains diffusion
controlled. The magnitude of the deviation is therefore governed not simply
by whether grain-boundary diffusion is present, but by the contrast between
grain-boundary and bulk transport and by the rate at which the grain-boundary
network evolves.

The kinetics are also sensitive to the grain-growth parameters. Increasing $k$,
corresponding to faster microstructural coarsening, shortens the period over which
the product layer retains a strong grain-boundary contribution. Conversely,
smaller $k$ values prolong the transient regime. An analogous effect 
is obtained by increasing the grain-growth exponent $m$, which corresponds
to more strongly retarded grain growth within the adopted formulation.
Thus, the duration of the non-parabolic regime is controlled
by the competition between reaction-layer thickening and microstructural coarsening.

These results demonstrate that a single power-law exponent cannot, in general,
be regarded as an intrinsic kinetic parameter of the reaction. The instantaneous
exponent continuously evolves as the diffusion coefficient changes. Nevertheless, the evolution can be conveniently interpreted in terms of three characteristic regimes.

\begin{figure*}
  \centering
    \includegraphics{"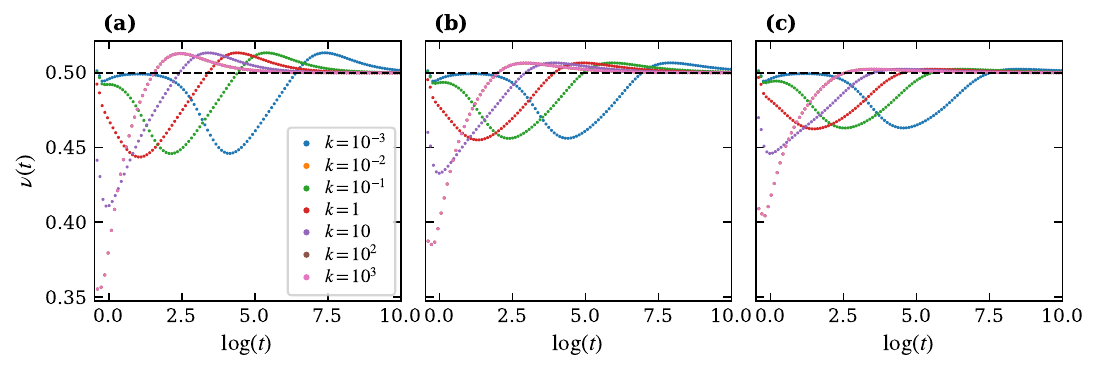"}
    \caption{Effect of the grain-growth on the instantaneous product-layer growth exponent for a relatively weak grain-boundary contribution, $\mathcal{D}_{GB}=1$.
	    The dependence of $\nu$ on $\log(t)$ is shown for different grain-growth
	    constants, $k$, for (\textit{a}) $m=2$, (\textit{b}) $m=3$, and (\textit{c}) $m=4$.}\label{fig3}
\end{figure*}

\begin{figure*}
  \centering
    \includegraphics{"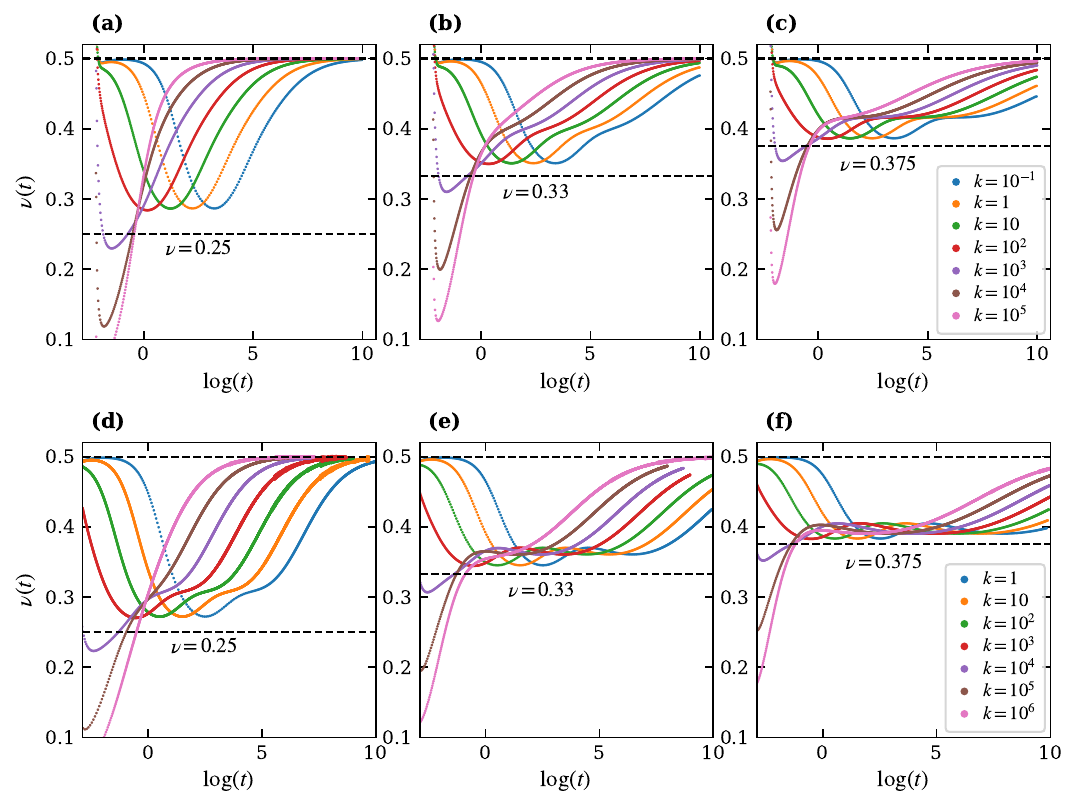"}
    \caption{Evolution of the instantaneous product-layer growth exponent for enhanced grain-boundary transport.
	    Results are presented for $\mathcal{D}_{GB} = 100$ (\textit{a--c}) and $\mathcal{D}_{GB} = 1000$ (\textit{d--f}) for different grain-growth 
	    constants, $k$, and grain growth exponents: (\textit{a, d}) $m =2$,
(\textit{b, e}) $m =3$,(\textit{c, f}) $m =4$.
Black dashed lines indicate the parabolic-growth exponent ($\nu = 0.5$) and the
corresponding expected growth exponents in asymptotic limit ($D_{B} \to 0$, $L_0 \to 0$).}\label{fig4}
\end{figure*}

\subsection{Transition between grain-boundary- and bulk-diffusion-controlled regimes}

\begin{figure}
  \centering
    \includegraphics{"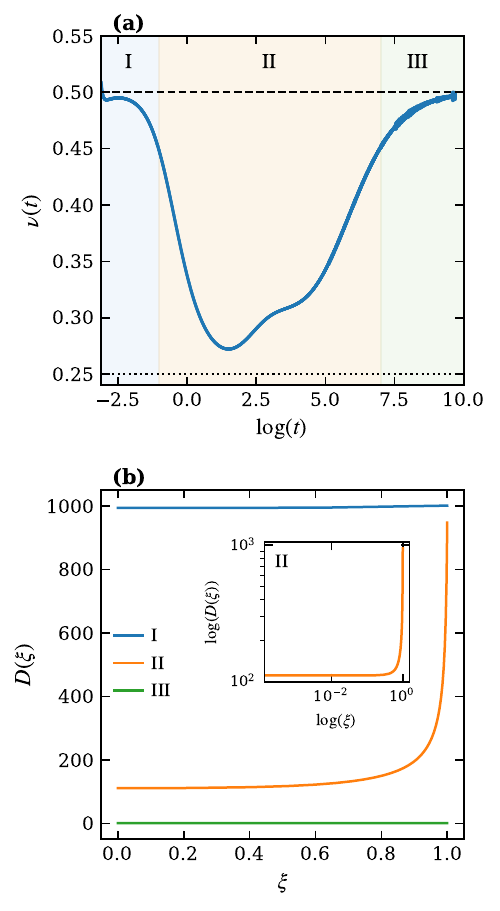"}
    \caption{Transition between grain-boundary- and bulk-diffusion-controlled product-layer growth. 
	    (\textit{a}) Instantaneous growth exponent, $\nu$, as a function of $\log(t)$ for $\mathcal{D}_{GB}=1000$ and $k=10$, showing three successive kinetic regimes: (I) initial grain-boundary-dominated parabolic growth, (II) transient sub-parabolic growth, and (III) late-stage bulk-diffusion-dominated parabolic growth.
	    (\textit{b}) Corresponding spatial distributions of the effective interdiffusion coefficient, $D(\xi)$, for representative times within the three regimes. }\label{fig5}
\end{figure}

For sufficiently large $\mathcal{D}_{\mathrm{GB}}$, three successive kinetic regimes
can be distinguished, as shown on Figure~\ref{fig5}. In the initial regime (I),
the product layer is dominated by grain-boundary transport. Because the newly
formed product has a small grain size, the grain-boundary volume fraction
is relatively high and the effective diffusivity is close to its maximum value.
The diffusivity is approximately uniform throughout the layer,
and the product-layer thickness follows an approximately parabolic
dependence with $\nu\simeq1/2$.

As the reaction proceeds, the microstructural ages of the material located
near the stationary interface and near the moving reaction front become 
increasingly different. Grain growth is therefore more advanced in the older 
portion of the product layer, while newly formed material at the reaction
front retains a fine-grained structure. The effective diffusivity consequently
develops a pronounced spatial gradient. This condition defines the intermediate
regime (II), in which the instantaneous growth exponent $\nu$ falls below $1/2$.

At sufficiently long times, grain coarsening reduces the grain-boundary 
contribution over most of the product layer. The effective diffusivity
consequently approaches the bulk diffusivity, $D\to D_B$, and becomes again
approximately uniform. The system then enters regime (III), characterized
by bulk-diffusion-controlled parabolic growth and $\nu\simeq1/2$.

These three regimes emerge naturally from a continuous evolution of the same
transport mechanism as the microstructure evolves. The transition from 
regime I to regime II reflects the development of spatially heterogeneous
microstructural age, whereas the transition from regime II to regime III 
reflects the progressive disappearance of the grain-boundary contribution
to the effective diffusivity.

The extent of regime II increases with increasing $\mathcal{D}_{\mathrm{GB}}$,
because a larger difference between grain-boundary and bulk diffusion amplifies
the effect of grain coarsening on transport. It also becomes more extended 
for slower grain growth, corresponding to smaller $k$ or larger $m$. Hence,
pronounced sub-parabolic behavior requires two conditions to be satisfied
simultaneously: the grain-boundary contribution must be sufficiently
large to influence the overall flux, and the associated microstructural
evolution must occur on a timescale comparable to that of reaction-layer growth.

An important consequence follows from the spatial distribution of the effective
diffusivity. During regime II, $D(\xi)$ does not follow the simple asymptotic
power-law dependence that would be obtained under the limiting assumptions 
of vanishing initial grain size and negligible bulk diffusion\cite{Nikiforov2025}.
Instead, the calculated diffusivity exhibits substantial curvature when plotted
on logarithmic coordinates. This indicates that the sub-parabolic
behavior does not require a specific analytical form of $D(\xi)$.
The essential condition is the presence of a sufficiently strong
spatial variation in the effective diffusivity across the growing layer.

This observation extends the interpretation of the present results beyond
grain growth alone. Any microstructural process capable of producing a comparable
spatial and temporal evolution of transport properties may generate analogous
deviations from parabolic kinetics. Examples include precipitation or dissolution 
of secondary phases, recrystallization, deformation-induced changes in defect density,
and the development of microcracks. The kinetic exponent below $1/2$ should therefore
be regarded as a signature of evolving transport properties rather than
as unique evidence for grain-boundary diffusion coupled to grain growth.

\subsection{Apparent power-law exponents and experimental interpretation}

Although the instantaneous exponent provides direct information about
the evolving kinetics, experimental studies commonly determine a single
exponent by fitting product-layer thickness measurements over a finite
time interval. To assess the implications of this procedure, the calculated
thickness curves were sampled over experimentally relevant time intervals
and subjected to a prescribed measurement uncertainty. The resulting datasets
were fitted using the conventional power-law representation,
Eq.~\ref{eq:log_power_law}.

\begin{figure}
  \centering
    \includegraphics{"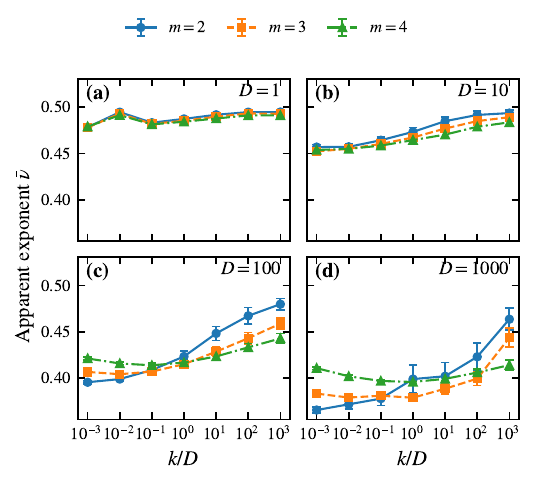"}
    \caption{Apparent product-layer growth exponent, $\bar{\nu}$, obtained by fitting the calculated layer-thickness data to a power law over finite observation intervals. Results are shown as a function of the ratio $k/\mathcal{D}_{GB}$ for $\mathcal{D}_{GB}= 1$ (\textit{a}), $10$ (\textit{b}), $100$ (\textit{c}), and $1000$ (\textit{d}), and for grain-growth exponents $m=2$, $3$, and $4$.}\label{fig6}
\end{figure}

Despite the continuous evolution of $\nu(t)$, the resulting thickness curves 
can be represented accurately by a power law over finite observation intervals
(Figure~\ref{fig6}). The apparent exponents fall within the range of
approximately 0.25–0.5, consistent with the range commonly associated with
sub-parabolic reaction-layer growth. Importantly, however, the apparent
exponent does not coincide directly with the instantaneous exponent associated
with a particular asymptotic grain-growth mode.

The dependence of $\bar{\nu}$ on the model parameters provides further
insight into this distinction. Increasing $\mathcal{D}_{\mathrm{GB}}$
generally decreases the apparent exponent because a larger grain-boundary
contribution produces a stronger and longer-lived modification of 
the effective diffusivity during grain coarsening. In contrast,
increasing $k$ increases the apparent exponent because faster
grain growth accelerates the transition toward bulk-diffusion-dominated
transport and therefore reduces the duration over which sub-parabolic
kinetics are observed.

The strong sensitivity of $\bar{\nu}$ to $k$ has an important experimental
implication. Although different grain-growth laws produce distinct instantaneous
kinetic behavior, their signatures can become difficult to distinguish
when the reaction is characterized solely by a fitted thickness exponent.
In particular, the apparent exponent cannot be used reliably to infer 
the grain-growth exponent $m$ without independent information concerning
the microstructural evolution and the relative magnitude of grain-boundary 
and bulk diffusion.

At the same time, the relatively small statistical variation of the fitted
exponent demonstrates why a continuously evolving kinetic process may
nevertheless appear experimentally to obey a well-defined power law.
The calculated standard deviation of the apparent exponent is below
approximately 5\% and is below 1\% for most of the considered cases.
Consequently, repeated measurements over the same finite time interval
can yield a highly reproducible exponent even though this exponent
represents an interval-averaged description of a non-stationary kinetic process.

The distinction between $\nu(t)$ and $\bar{\nu}$ is therefore essential
when interpreting experimental reaction kinetics. A measured exponent
below $1/2$ does not necessarily imply that the system possesses
a single asymptotic sub-parabolic growth law. Instead, it may represent
the effective slope of a transient process in which the dominant
contribution to transport changes continuously from grain-boundary to bulk diffusion.

Overall, the calculations demonstrate that grain growth can transform
an initially grain-boundary-dominated diffusion process into 
a bulk-diffusion-dominated one through a continuous evolution of 
the local effective diffusivity. The resulting product-layer growth kinetics
naturally pass through an intermediate sub-parabolic regime. 
The magnitude and duration of this regime are determined
by the competition between grain-boundary and bulk diffusion 
and the kinetics of microstructural coarsening. The framework
thus provides a mechanistic basis for interpreting non-parabolic growth
in solid-state diffusion couples without requiring the introduction 
of an independent, time-invariant kinetic exponent.


\section{Conclusions}
A numerical model was developed to describe diffusion-controlled growth 
of a product layer in a solid-state diffusion couple while explicitly 
accounting for the evolution of the grain structure. 
In contrast to models employing a spatially uniform, time-independent diffusivity,
the present formulation accounts for the fact that different portions
of a continuously growing product layer have different microstructural 
ages and therefore different grain sizes and effective diffusivities.
Both grain-boundary and bulk diffusion were retained, allowing
the transition between the two transport contributions to emerge
naturally from the coupled evolution of the product layer and its microstructure.

The principal conclusions are as follows:
\begin{enumerate}
\item The numerical formulation reproduces the classical parabolic growth law
	in the absence of grain growth. When the effective interdiffusion
	coefficient is constant, the calculated product-layer thickness
	agrees with the analytical diffusion-controlled solution after
	accounting for the finite initial thickness used in the numerical formulation.
\item Grain growth produces a transient departure from parabolic product-layer
	growth. The instantaneous growth exponent is not generally constant
	when the product-layer diffusivity evolves as a consequence
	of microstructural coarsening. The strongest deviations from $\nu=1/2$
	occurs when the grain-boundary contribution to transport
	is sufficiently large and the grain-growth timescale is comparable
	to that of product-layer thickening.
\item Three successive kinetic regimes can be identified. The reaction 
	initially exhibits approximately parabolic growth dominated
	by grain-boundary diffusion. This regime is followed by a transient
	sub-parabolic regime associated with strong spatial variation
	of the effective diffusivity across the product layer. At longer times,
	grain coarsening reduces the grain-boundary contribution and
	the system approaches a second approximately parabolic regime
	dominated by bulk diffusion. Thus, sub-parabolic growth represents
	a transient consequence of evolving transport properties.
\item The extent of the sub-parabolic regime is controlled by the competition
	between grain growth and diffusion. Increasing the relative
	grain-boundary diffusivity increases the magnitude of the deviation
	from parabolic behavior, whereas slower grain coarsening extends
	the duration of the transient regime.
\item Despite the continuously evolving $\nu(t)$, product layer thickness
	can be represented sufficiently accurately by a power law over
	a finite experimental time interval, yielding a reproducible
	apparent exponent $\bar{\nu}$.
	The apparent exponent depends on the grain-boundary contribution
	and the kinetics of grain growth and therefore cannot, by itself,
	be used to uniquely determine the underlying grain-growth law.
\end{enumerate}

Overall, the results demonstrate that the commonly used power-law description
of product-layer growth can conceal a substantially more complex transient
transport behavior. In a polycrystalline product layer, the continuous
formation of new material and simultaneous grain coarsening lead to
an effective diffusivity that evolves both spatially and temporally
as a consequence of the local microstructural history. Incorporating 
this history into the diffusion model provides a physically consistent
description of the transition from grain-boundary- to bulk-diffusion-controlled
growth and offers a mechanistic interpretation of experimentally
observed deviations from parabolic kinetics. The results further
indicate that sub-parabolic growth should not be regarded as unique
evidence of grain-boundary diffusion coupled with grain growth. Rather,
the essential condition is a sufficiently pronounced evolution of 
the effective diffusivity during reaction. Consequently, other
microstructural processes that alter the transport properties
of the product layer may give rise to analogous kinetic behavior.
The present framework therefore provides a basis for relating
microstructural evolution to reaction-layer growth and may be
extended to the analysis and control of interlayer properties
under conditions of evolving microstructure.










\bibliographystyle{elsarticle-num}


\bibliography{references}



\end{document}